\documentstyle[preprint,aps]{revtex}  
\begin{document} 
\draft  
\title{Charge quantization in the largest 
leptoquark-bilepton chiral electroweak scheme}    
\author{A. Doff and F. Pisano}    
\address{Departamento de F\'\i sica, Universidade Federal do Paran\'a, 
81531-990 Curitiba, PR, Brazil}  
\date{\today}  
\maketitle 
\begin{abstract} 
The uniqueness of the hypercharge assignments in the 
three fermion families    
leptoquark-bilepton SU(3)$_C\times$SU(4)$_L\times$U(1)$_N$ model   
is established. Although the gauge group contains an explicit  
U(1) factor, freedom from triangle anomalies combined with  
the requirement of nonvanishing charged fermion masses uniquely  
fix the electric charges of all fermions independently of the     
neutrinos being massless or not. The electric-charge    
quantization, family replication, and the existence of three 
colors are interwoven.      
%%%%%%%%%%%%%%%%%%%%%%%%%%%%%%%%%%%%%%%%%%%%%%%%%%%%%%%% 
%%Electric-charge quantization does not depend on 
%%the neutrino masses 
%%and is realized either the neutral fermions 
%%are massless Weyl, Dirac or Majorana massive fields.        
%%%%%%%%%%%%%%%%%%%%%%%%%%%%%%%%%%%%%%%%%%%%%%%%%%%%%%%%
\end{abstract} 
\bigskip
\pacs{PACS numbers: 11.30.Er, 11.30.Hv, 12.15.Cc}     
%%\bigskip 
\newpage  
So far, the standard model                
of electroweak and strong interactions~\cite{sm}         
has been quite successful in its compatibility with almost  
all available experimental data~\cite{kamiokalsnd,pdg98}.   
It nevertheless leaves  
some fundamental theoretical questions unexplained.      
In the standard model, each family of fermions is  
anomaly-free and this is true as well for grand unified models,   
supersymmetric extensions, except the supersymmetric  
preon model~\cite{preon},     
technicolor, superstring theories, 
and most compositeness scenarios where the number of families remains      
completely unrestricted on theoretical grounds. The chiral anomaly   
is cancelled between quarks and leptons in each family and the     
indetermination about the inter-relation between families    
constitute the so-called flavor question. At present, we know of  
three families, but the standard model does not explain why this  
number has to be three, even so the number of neutrino flavors  
within the electroweak standard model is   
$N_\nu=3.00\pm0.09$ and the experimental determination of this  
number is model dependent~\cite{lep}.    
Some very fundamental aspects of the standard model such as the 
flavor question might be understood by embedding the three-family 
version in a Yang-Mills theory with the gauge semisimple 
group 
$$ 
G_0\equiv {\rm SU(3)}_C\otimes {G_W}\otimes {\rm U(1)}_{L+R}  
$$  
just enlarging the SU(2)$_L$ weak isospin group to   
$G_W={\rm SU(3)}_L$ (331 model~\cite{pp92,Framp92,footpp93})       
which is the minimal gauge group that at the leptonic  
level admits charged fermions and their antiparticles as members of   
the same multiplet~\cite{adler89}. The key predictions of the  
$G_0$ alternative models are leptoquark fermions with electric  
charges $\pm 5/3$ and $\mp4/3$ and bilepton gauge   
bosons~\cite{cuypers96} with lepton number $L=\pm 2$.   
The leptoquark fermions are   
color-triplet particles which possess barion and lepton numbers.     
Another interesting feature of these   
leptoquark-bilepton chiral models is that the weak mixing angle   
of the standard model has an upper limit. Therefore, it is possible   
to compute an upper bound to the mass scale of the    
$G_W={\rm SU(3)}_L$ breaking of about $1.7$ TeV~\cite{ng94}.     
\par%%%talvez%%% 
Considering the lightest particles of the model as the sector in  
which a symmetry is manifested, the lepton sector could be the   
part of the model determining new approximate symmetries. In fact,  
if right-handed neutrinos are introduced, there arises a more   
interesting possibility of having $\nu_l$, $l=e,\mu,\tau$, and the 
charge conjugate fields $\nu^c_l$, $l^c$ in the same multiplet for  
each family flavor.  
Model building in that direction, if each family of fermions is  
treated separately, culminates with the highest symmetry   
$G_W={\rm SU(4)}_L$ to be considered in the electroweak 
sector (341 model~\cite{voloshin,pitran93,foot94,pp95}).     
In the 331 and 341 leptoquark-bilepton models    
the number of families must be divisible by the number  
of color degrees of freedom in order to cancel anomalies.     
This novel method of anomaly  
cancellation requires that at least one family transforms  
differently from the others, thus breaking generation universality.  
To accommodate the replication of three fermion families,       
the number of families, number of colors, and fractional electric  
charge values become related~\cite{Framp92,framp93,pisano96}.     
Having established that connection the flavor question is solved   
with a relation between the strong and electroweak parts of the  
model which does not exist in the context of the standard   
model.     
In the minimal standard model there is a remarkable failure  
concerning the connection among family replication and the   
electric charge quantization~\cite{foot91}.      
In fact, the charge quantization is realized only within each  
family~\cite{footetal,babu89,babu90,rudazetal,zralek}. 
Nevertheless, 
taking the three families together the effect of dequantization  
occurs~\cite{footetal,babu89,zralek,pilaftsis}.      
The possibility of charge quantization with three families in the   
minimal 331 model was shown recently~\cite{ravinez}.    
\par 
In the 341 model, the electric charge operator is embedded in 
the neutral generators of the SU(4)$\otimes$U(1) group   
\begin{equation}  
{\cal Q} = \Lambda_3 + \xi\Lambda_8 + \zeta\Lambda_{15} 
+ \varepsilon N
\label{carga}
\end{equation}
with the embedding parameters $\xi=-1/\sqrt 3$, $\zeta=-2\sqrt 6/3$   
and $N$ is the new U(1) charge. 
The neutral generators of SU(4)$_L$ are 
$$
\Lambda_3=\frac{\lambda_3}{2}=\frac{1}{2}{\rm diag}(1,-1,0,0),   
$$
$$  
\Lambda_8=\frac{\lambda_8}{2}=\frac{1}{2\sqrt 3}{\rm diag}(1,1,-2,0),   
$$
$$ 
\Lambda_{15}=\frac{\lambda_{15}}{2}=\frac{1}{2\sqrt 6}{\rm diag}(1,1,1,-3).   
$$
The model treats the color singlet leptons democratically in each 
of the three families  
\begin{equation}  
f_{iL} = 
\left (
\begin{array}{c}
\nu_i \\ 
l_i \\ 
\nu^c_i \\
l^c_i
\end{array}
\right )_L 
\sim ({\bf 1},{\bf 4},N_i)
\label{leptons}
\end{equation}
where $i=1,2,3$ is a flavor index and $\nu^c_i$, $l^c_i$ denote    
charge conjugated fields. There are no leptonic flavor singlets   
because right-handed charged leptons are not independent degrees of  
freedom and can be obtained through charge conjugation of the 
fields contained in the multiplet $f_{iL}$. The quarks have the  
attributions  
\begin{equation} 
Q_{1L} = 
\left (
\begin{array}{c}  
u_1 \\ 
d_1 \\ 
u^\prime \\ 
J
\end{array} 
\right )_L 
\sim ({\bf 3},{\bf 4},N_{Q_{1L}})  
\label{quarkum}
\end{equation}
\begin{equation} 
Q_{\alpha L} = 
\left (
\begin{array}{c}
j_\alpha \\ 
d^\prime_\alpha \\ 
u_\alpha \\ 
d_\alpha
\end{array}
\right )_L 
\sim ({\bf 3},{\bf 4^*},N_{Q_{\alpha L}})  
\label{quark2e3}
\end{equation}
and the associated right-handed projections transform as singlets  
under SU(4)  
\begin{mathletters}
\begin{eqnarray}
u_{1R} & \sim & ({\bf 3},{\bf 1},N_{u_{1R}})  
\label{qr1}
\\ 
d_{1R} & \sim & ({\bf 3},{\bf 1},N_{d_{1R}})  
\label{qr2} 
\\ 
u^\prime_R & \sim & ({\bf 3},{\bf 1},N_{{u^\prime}_R}) 
\label{qr3} 
\\ 
J_R & \sim & ({\bf 3},{\bf 1},N_{J_R})  
\label{qr4} 
\\ 
j_{\alpha R} & \sim & ({\bf 3},{\bf 1},N_{j_{\alpha R}})  
\label{qr5}
\\ 
%%%%nome velho: quarkright%%%%%%%%%%%%%  
d^\prime_{\alpha R} & \sim & ({\bf 3},{\bf 1},N_{d^\prime_{\alpha R}})   
\label{qr6} 
\\ 
u_{\alpha R} & \sim &  ({\bf 3},{\bf 1},N_{u_{\alpha R}})  
\label{qr7} 
\\ 
d_{\alpha R} & \sim & ({\bf 3},{\bf 1},N_{d_{\alpha R}})  
\label{qr8}
\end{eqnarray}
\end{mathletters}
where $\alpha=2,3$.   
%%%\par  
The 341 original symmetry is broken and quark masses are generated  
by the following Higgs multiplets 
\begin{eqnarray} 
\eta & \sim & ({\bf 1},{\bf 4},N_\eta), 
\nonumber \\ 
\rho & \sim & ({\bf 1},{\bf 4},N_\rho), 
\\
\label{higgstri}
\chi & \sim & ({\bf 1},{\bf 4},N_\chi).      
\nonumber 
\end{eqnarray} 
The lepton mass term transforms as   
$\overline{(f_L)^c}f_L \sim ({\bf 1}, {\bf 6}_A\oplus {\bf 10}_S)$.  
In order to obtain massive charged leptons it is  necessary to 
introduce the multiplet     
\begin{equation}  
H^* = \left ( 
\begin{array}{cccc}
H_1^0 & H_1^+ & H_2^0 & H_2^- \\ 
H_1^+ & H_1^{++} & H_3^+ & H_3^0 \\ 
H_2^0 & H_3^+ & H_4^0 & H_4^- \\   
H_2^- & H_3^0 & H_4^- & H_2^{--} 
\end{array} 
\right ) \sim ({\bf 1},{\bf 10}^*_S,N_{H^*})
\end{equation} 
because the ${\bf 6}^*_A$ will leave one lepton massless and two  
others degenerate for three generations.  
Therefore a vacuum expectation value of the decuplet is  
needed to produce a realistic leptonic mass matrix.    
Moreover, in order to   
avoid mixing among primed and unprimed quarks we introduce  
another multiplet $\eta^\prime$ transforming as $\eta$ but with   
different vacuum expectation value. Notice that the introduction   
of the (anti)decuplet $H^*$ is not essential for the symmetry  
breaking. In fact, the $341$ gauge symmetry breaks to   
SU(3)$_C\times$U(1)$_{\cal Q}$ if the vacuum  
structures           
$\langle\eta\rangle = (v_\eta,0,0,0)$,    
$\langle\rho\rangle = (0,v_\rho,0,0)$,  
$\langle\chi\rangle = (0,0,0,v_\chi)$, 
$\langle\eta^\prime\rangle = (0,0,v_{\eta^\prime},0)$,      
are realized.       
At tree level the charged leptons get a mass but neutrinos remain   
massless if $\langle H_3^0\rangle\neq 0$,  
$\langle H^0_{1,2,4}\rangle = 0$. 
\par 
The electric charge operator is defined as the linear combination  
\begin{equation} 
{\cal Q} = T_3 + \frac{Y}{2}
\label{opcarga} 
\end{equation}
which annihilates the vacuum. Consequently,   
\begin{equation}  
N_\eta = N_{\eta^\prime} = N_{H^*} = 0, 
\quad \varepsilon =\frac{1}{N_\rho},
\quad N_\chi = - N_\rho, 
\label{vinculos}
\end{equation}   
so that the hypercharge is the mixture   
\begin{equation}  
\frac{Y}{2} = T_8 + T_{15} + \frac{N}{N_\rho}  
\label{hyper}
\end{equation} 
with the components       
$$  
T_8 = \xi\Lambda_8 = -\frac{1}{2\sqrt 3}\lambda_8, 
$$        
$$
T_{15} = \zeta\Lambda_{15} = -\frac{\sqrt 6}{3}\lambda_{15},   
$$
and $T_3=\Lambda_3=\lambda_3/2$.  
The most general Yukawa interactions in terms of weak eigenstates 
are 
\begin{eqnarray} 
-{\cal L}_Y = & \frac{1}{2} & G_{ij}\overline{(f_{iL})^c}H^*f_{jL}  
\nonumber \\ 
& + & F_{1k}\bar Q_{1L}u_{kR}\eta + 
F_{\alpha k}\bar Q_{\alpha L}u_{kR}\rho^* 
\nonumber \\ 
& + & F^\prime_{1k}\bar Q_{1L}d_{kR}\rho + F^\prime_{\alpha k} 
\bar Q_{\alpha L}d_{kR}\eta^* 
\nonumber \\ 
& + & h_1\bar Q_{1L}u^\prime_R\eta^\prime + 
h_{\alpha\beta}\bar Q_{\alpha L}d^\prime_{\beta R}\eta^{\prime *}  
\nonumber \\  
& + & \Gamma_1\bar Q_{1L}J_R\chi + 
\Gamma_{\alpha\beta}\bar Q_{\alpha L}j_{\beta R}\chi^* + {\rm H.c.} 
\label{yuk}
\end{eqnarray}
where $i,j=e,\mu,\tau$; $k=1,2,3$; and $\alpha,\beta =2,3$. These    
couplings automatically contain a Peccei-Quinn symmetry~\cite{pq77}     
which can also be extended to the Higgs potential, solving the strong  
$CP$ problem~\cite{pal95}.    
%%%\par  
The U(1)$_N$ gauge invariance of the Yukawa leptonic term gives    
three classical constraints 
\begin{equation}  
N_i = 0, \quad i=e,\mu,\tau   
\label{lepcon}
\end{equation}
while for the quark flavors the set of classical constraints is  
\begin{mathletters} 
\begin{eqnarray}  
N_{Q_{1L}}-N_{u_{kR}} & = & N_\eta, 
\label{clco1} 
\\
N_{Q_{\alpha L}}-N_{d_{kR}} & = & N_{\eta^*}, 
\label{clco2} 
\\  
N_{Q_{1L}}-N_{d_{kR}} & = & N_\rho, 
\label{clco3} 
\\ 
N_{Q_{\alpha L}}-N_{u_{kR}} & = & N_{\rho^*}, 
\label{clco4}
\\  
N_{Q_{1L}}-N_{u^\prime_R} & = & N_{\eta^\prime},   
\label{clco5} %%%%%%%nome velho: clacon%%%%%%% 
\\
N_{Q_{\alpha L}}-N_{d^\prime_{\alpha R}} & = & N_{\eta^{\prime *}}, 
\label{clco6} 
\\ 
N_{Q_{1L}}-N_{J_R} & = & N_\chi, 
\label{clco7} 
\\ 
N_{Q_{\alpha L}}-N_{j_{\alpha L}} & = & N_{\chi^*},   
\label{clco8} 
\end{eqnarray}
\end{mathletters} 
where $N_\eta=-N_{\eta^*}$, $N_{\eta^\prime}=-N_{\eta^{\prime *}}$,  
$N_{\rho}=-N_{\rho^*}$, and $N_\chi=-N_{\chi^*}$. These conditions   
imply  
\begin{eqnarray}   
N_{Q_{1L}}+N_{Q_{2L}} & = & N_{u_R}+N_{s_R},   
\nonumber \\
N_{Q_{1L}}+N_{Q_{2L}} & = & N_{d_R}+N_{c_R},  
\\ 
\label{cc1}
N_{Q_{1L}}+N_{Q_{2L}} & = & N_{j_{2R}}+N_{J_R}, 
\nonumber  
\end{eqnarray}
for the first and second families and  
\begin{eqnarray}  
N_{Q_{2L}}-N_{Q_{3L}} & = & N_{j_{2R}}-N_{j_{3R}},  
\nonumber \\
N_{Q_{2L}}-N_{Q_{3L}} & = & N_{c_R}-N_{t_R}, 
\\ 
\label{cc2}
N_{Q_{2L}}-N_{Q_{3L}} & = & N_{s_R}-N_{b_R}, 
\nonumber
\end{eqnarray} 
which relates the second and third families. Therefore, we obtain  
the following two conditions, 
\begin{equation} 
N_{Q_{2L}}=N_{Q_{3L}}\equiv N_{Q_{\alpha L}}, \quad \alpha=2,3
\label{cond1}
\end{equation}  
and 
\begin{equation}  
N_{j_{2R}}=N_{j_{3R}}\equiv N_{j_{\alpha R}}.   
\label{cond2}
\end{equation}
\par   
To consider the quantum constraints, let us set the following   
notation  
$$
N_{u_{1R}}=N_{u_{2R}}=N_{u_{3R}}\equiv N_{U_R}, 
$$
$$
N_{d_{1R}}=N_{d_{2R}}=N_{d_{3R}}\equiv N_{D_R}.     
$$
for the up- and down-like standard flavors.        
It will be sufficient to look at the anomalies which contain U(1)$_N$   
factors,   
\\
\begin{mathletters}  
\centerline{Tr[SU(3)$_C$]$^2$[U(1)$_N$]=0:}  
\begin{equation}  
3(N_{Q_{1L}}+2N_{Q_{\alpha L}})-3(N_{U_R}+N_{D_R})-N_{J_R}
-2N_{j_{\alpha R}}-N_{u^\prime_R}-2N_{d^\prime_{\alpha R}} = 0, 
\label{anum}
\end{equation} 
\centerline{Tr[SU(4)$_L$]$^2$[U(1)$_N$]=0:}    
\begin{equation}  
3(N_{Q_{1L}}+2N_{Q_{\alpha L}})+\sum_iN_i = 0,  
\label{andois} 
\end{equation}  
\centerline{Tr[U(1)$_N$]$^3$=0:}   
\begin{equation}  
3(N^3_{Q_{1L}}+2N^3_{Q_{\alpha L}})-3(N^3_{U_R}+N^3_{D_R})   
-N^3_{J_R}-2N^3_{j_{\alpha R}}-N^3_{u^\prime_R}
-2N^3_{d^\prime_{\alpha R}}+\sum_iN^3_i = 0 
\label{antres}
\end{equation}
\centerline{Tr[graviton]$^2$[U(1)$_N$]=0:}   
\begin{equation}  
3(N_{Q_{1L}}+2N_{Q_{\alpha L}})-3(N_{U_R}+N_{D_R})-N_{J_R}
-2N_{j_{\alpha R}}-N_{u^\prime_R}-2N_{d^\prime_{\alpha R}}  
+\sum_iN_i=0
\label{anqua}
\end{equation} 
\end{mathletters}  
where the last condition arises from a triangle graph with two   
external gravitons. Whatever the correct   
quantum gravity theory is, the mixed gauge-gravitational   
anomaly~\cite{delbourgo} must be canceled to ensure     
the general covariance of the theory.   
The other non-trivial anomaly is [SU(4)$_L$]$^3$ which also cancels   
between the families if the number of fermion families 
coincides with the number of SU(3)$_C$ color degrees of 
freedom~\cite{pisano96}. Through the leptonic classical conditions 
of Eq. (\ref{lepcon}) we have  
$$ 
\sum_iN_i = 0 
$$  
over the leptonic families. Therefore, the mixed gauge-gravitational  
constraint coincides with that of the [SU(3)$_C$]$^2$[U(1)$_N$]   
anomaly involving the color gauge bosons. As such there are        
essentially only three independent quantum constraints. 
\par 
The U(1)$_N$ gauge invariance of the quark Yukawa couplings    
gives the explicit classical constraints   
\begin{mathletters} 
\begin{eqnarray}  
N_{U_R} & = & N_{Q_{\alpha L}} + N_\rho, 
\label{tuum}
\\
N_{D_R} & = & N_{Q_{\alpha L}},  
\label{tudois}
\\ 
N_{J_R} & = & N_{Q_{\alpha L}} - N_\rho,  
\label{tutres}
\\
N_{u^\prime_R} & = & N_{Q_{1L}}, 
\label{tuqua}
\\  
N_{d^\prime_{\alpha R}} & = & N_{Q_{\alpha L}},    
\label{tudo} 
\end{eqnarray}  
\end{mathletters} 
and also, by the quantum constraints, there are the additional  
conditions          
\begin{mathletters}  
\begin{equation}  
N_{Q_{1L}} = N_{Q_{\alpha L}} + N_\rho, 
\label{claqua1}
\end{equation}
\begin{equation}
N_{J} = N_{Q_{\alpha L}} + 2N_\rho.  
\label{claqua2}
\end{equation} 
\end{mathletters} 
Let us take the condition in Eq. (\ref{claqua1}) and the   
[SU(4)$_L$]$^2$[U(1)$_N$] anomaly constraint  
\begin{equation}  
N_{Q_{1L}} + 2N_{Q_{\alpha L}} = 0
\label{3L1}
\end{equation}
which, in turn, may be related to give 
\begin{equation}  
N_{Q_{\alpha L}} = -\frac{1}{3} N_\rho  
\label{q1q} 
\end{equation}
which establish the U(1)$_N$ quark charge relations in units of   
$N_\rho$ 
\begin{mathletters}  
\begin{eqnarray} 
N_U & = & \frac{2}{3}N_\rho, 
\\
\label{num}
N_D & = &  -\frac{1}{3}N_\rho, 
\\ 
\label{ndois}
N_{Q_{1L}} & = & \frac{2}{3}N_\rho, 
\\
\label{ntres}
N_{Q_{\alpha L}} & = & -\frac{1}{3}N_\rho, 
\\ 
\label{nqua}
N_{J} & = & \frac{5}{3}N_\rho,  
\\
\label{ncinco}
N_{j_{\alpha L}} & = & -\frac{4}{3}N_\rho, 
\label{nseis}
\end{eqnarray}
and 
\begin{eqnarray}  
N_{u^\prime_R} & = & N_{Q_{1L}} = \frac{2}{3}N_\rho,  
\label{nsete} 
\\
N_{d^\prime_{\alpha R}} & = & N_{Q_{\alpha L}} = -\frac{1}{3}N_\rho. 
\label{noito}
\end{eqnarray}
\end{mathletters} 
These $N$ charges for all quark flavors, together with the lepton 
ones in Eq. (\ref{lepcon}), allow to find the electric charges   
of all fermions in the 341 model by using the general expression  
of Eq. (\ref{opcarga}) that is, 
 \begin{eqnarray}  
{\cal Q}_{\nu_i} & = & 0,   
\nonumber \\ 
{\cal Q}_i & = & -1, 
\nonumber \\ 
{\cal Q}_U & = & \frac{2}{3}, 
\nonumber \\ 
{\cal Q}_D & = & -\frac{1}{3},  
\nonumber \\ 
{\cal Q}_{u_1} & = & \frac{2}{3}, 
\nonumber \\ 
{\cal Q}_{d^\prime_\alpha} & = & -\frac{1}{3}, 
\nonumber \\ 
{\cal Q}_{J} & = & \frac{5}{3}, 
\nonumber \\ 
{\cal Q}_{j_{\alpha}} & = & -\frac{4}{3}.  
\nonumber  
\end{eqnarray} 
\par  
Summarizing, following the approach of the electric charge 
quantization in models that contain an explicit U(1) factor in 
the semisimple gauge group, we have showen that the quantization    
of electric charge occurs in the largest leptoquark-bilepton  
chiral gauge extension when the three families are    
taken together even for      
massless neutrinos. In the minimal 341 model the neutrinos remain   
massless since there is a global symmetry which prevents them  
from getting a mass. This symmetry implies the conservation of the  
quantum number ${\cal F}=L+B$, where         
$L=L_e+L_\mu+L_\tau$ is the total lepton number and $B$ is the baryon    
number~\cite{pleton93}. If we allow this symmetry to be explicitly   
broken, Majorana neutrinos arise in the model if     
$\langle H^0_{1,4}\rangle\neq 0$ which conserves the structure of  
the leptonic sector in Eq. (\ref{yuk}). Also, if we add right-handed   
neutrino singlets, Dirac mass terms      
$\bar f_{iL}\eta\nu_{iR}$        
arise in the Yukawa couplings, but the gauge invariance     
of these terms implies   
$N_{\nu_{i}}=0$. Then the charge quantization is unavoidable and  
does not depends on the nature of the neutral fields. Moreover,        
family replication, charge quantization, 
the existence of three colors and absence of massless charged   
fermions are interconnected in the minimal 331 leptoquark-bilepton    
model and its largest extension.      
\acknowledgments  
We wish to thank S. Shelly Sharma for helpful discussions.    
%%%%%%%%%%%%%%%%%%%%%%%%%%%%%%%%%%%%%%%%%%%%%%%%%%%%%%%%%%%%%%%%%%%%

\end{document}